\documentclass[amsmath,amssymb,aps, prb,
showkeys,showpacs, superscriptaddress, floatfix, nofootinbib] {revtex4-2}
\usepackage{graphicx}
\usepackage{color}
\usepackage{bm}
\usepackage{hyperref}
\usepackage{dcolumn}
\usepackage{slashed}
\allowdisplaybreaks

\begin{document}
\title{Entanglement entropy in critical quantum spin chains with boundaries and defects}
\author{Ananda Roy}
\email{ananda.roy@physics.rutgers.edu}
\affiliation{Department of Physics and Astronomy, Rutgers University, Piscataway, NJ 08854-8019 USA}
\author{Hubert Saleur}
\affiliation{Institut de Physique Th\'eorique, Paris Saclay University, CEA, CNRS, F-91191 Gif-sur-Yvette}
\affiliation{Department of Physics and Astronomy, University of Southern California, Los Angeles, CA 90089-0484, USA}

\begin{abstract}
Entanglement entropy~(EE) in critical quantum spin chains described by 1+1D conformal field theories contains signatures of the universal characteristics of the field theory. Boundaries and defects in the spin chain give rise to universal contributions in the EE. In this work, we analyze these universal contributions for the critical Ising and XXZ spin chains for different conformal boundary conditions and defects. For the spin chains with boundaries, we use the boundary states for the corresponding continuum theories to compute the subleading contribution to the EE analytically and provide supporting numerical computation for the spin chains. Subsequently, we analyze the behavior of EE in the presence of conformal defects for the two spin chains and describe the change in both the leading logarithmic and subleading terms in the EE.
\end{abstract}
\maketitle 

\section{Introduction}
Entanglement, one of the quintessential properties of quantum mechanics, plays a central role in the development of long-range correlations in quantum critical phenomena. Thus, quantification of the entanglement in a quantum-critical system provides a way to characterize the universal properties of the critical point. The von-Neumann entropy of a subsystem serves as a natural candidate to perform this task. For zero-temperature ground-states of 1+1D quantum-critical systems described by conformal field theories~(CFTs), the von-Neumann entropy~[equivalently in this case, entanglement entropy~(EE)] for a subsystem exhibits universal logarithmic scaling with the subsystem size~\cite{Holzhey1994, Calabrese2004}. The coefficient of this scaling determines a fundamental  property of the bulk CFT:  the central charge, which quantifies, crudely speaking, the number of long-wavelength degrees of freedom. The aforementioned scaling, together with strong subaddivity property of entropy~\cite{Casini2004} and Lorentz invariance, leads to an alternate proof~\cite{Casini2006} of the celebrated $c$-theorem~\cite{Zamolodchikov1986} in 1+1 dimensions. At the same time, the scaling of EE in these gapless systems~\cite{Pollmann2009} and their gapped counterparts~\cite{Hastings2007} lies at the heart of the success of numerical techniques like density matrix renormalization group~(DMRG)~\cite{White1992, Schollwock2011} in simulating 1+1D quantum systems. 

Given the widespread success of EE in characterizing bulk properties of quantum-critical points, it is natural to ask if EE also captures signatures of boundaries and defects in gapless conformal-invariant systems. Consider CFTs on finite systems with conformal boundary conditions. For these systems, the EE receives a universal, subleading, boundary-dependent contribution, the so-called `boundary entropy'~\cite{Affleck1991,Calabrese2009}. The latter, related to the `ground-state degeneracy' of the system, plays a central role in a wide-range of problems both in condensed matter physics~\cite{Affleck1995, Saleur1998} and in string theory~\cite{Gaberdiel2000}. The boundary contribution in the EE is a valuable diagnostic for identifying the different boundary fixed points of a given  CFT~\cite{Affleck2009, Cardy2016, Roy2020a}. 

Conformal defects or interfaces comprise the more general setting. In the simplest case, a CFT, instead of being terminated with vacuum, is glued to another CFT~\cite{Bachas2001}. In general, the two CFTs do not have to be the same and have not even the same central charges~\cite{Quella:2002ct}. Unlike the case of boundaries, conformal defects can affect not just the subleading terms in the EE, but also the leading logarithmic scaling~\cite{Sakai2008, Eisler2010, Peschel2012e}. 
Of particular interest are the~(perfectly-transmissive) topological defects which can glue together CFTs of identical central charges~\cite{Petkova2000, Bachas2001, Frohlich2004, Frohlich2006,Aasen2016}. These defects commute with the generators of conformal transformations and thus, can be deformed without affecting the values of the correlation functions as long as they are not taken across field insertions~(hence the moniker topological). They reflect the internal symmetries of the CFT and relate the order-disorder dualities of the CFT to the high-low temperature dualities of the corresponding off-critical model~\cite{Frohlich2004, Krammers1941,Savit1980}. They also play an important role in the study of anyonic chains and in the correspondence between CFTs and three-dimensional topological field theories~\cite{Buican2017}. For these topological defects, the EE receives nontrivial contributions due to zero-energy modes of the defect Hamiltonian~\cite{Klich2017, Roy2021a}, which in turn can be used to identify the different defects~\cite{Roy2021a}. Note that although the EE of a subsystem in the presence of a defect contains information about the latter~\cite{Sakai2008, Brehm2015, Gutperle2015}, the exact behavior of the EE depends on the geometric arrangement of the subsystem with respect to the defect. 

The goal of this work is to describe the behavior of the EE for the Ising and the free, compactified boson CFTs in the presence of boundaries and defects. First, we compute the boundary entropy for these two models for free~(Neumann) and fixed~(Dirichlet) boundary conditions for the continuum theory using analytical techniques. This is done by computing the corresponding boundary states~\cite{Ishibashi1988, Cardy1989}. We compare these analytical predictions with numerical computations of the EE for suitable lattice regularizations. Subsequently, we investigate EEs in the two CFTs in the presence of defects when the defect is located precisely at the interface between the subsystem and the rest. For the Ising model, we consider the energy and duality defects. The computation of EE is performed by mapping the defect Hamiltonian to a free-fermion Hamiltonian and computing the ground-state correlation function. Finally, we analyze the EE across a conformal interface of two free, compactified boson CFTs. The numerical computation is done using DMRG for two coupled XXZ chains with different anisotropies. 

\section{Entanglement entropy in CFTs with boundaries}
\label{sec:EE_BCFT}
\begin{figure}
\centering
\includegraphics[width = 0.7\textwidth]{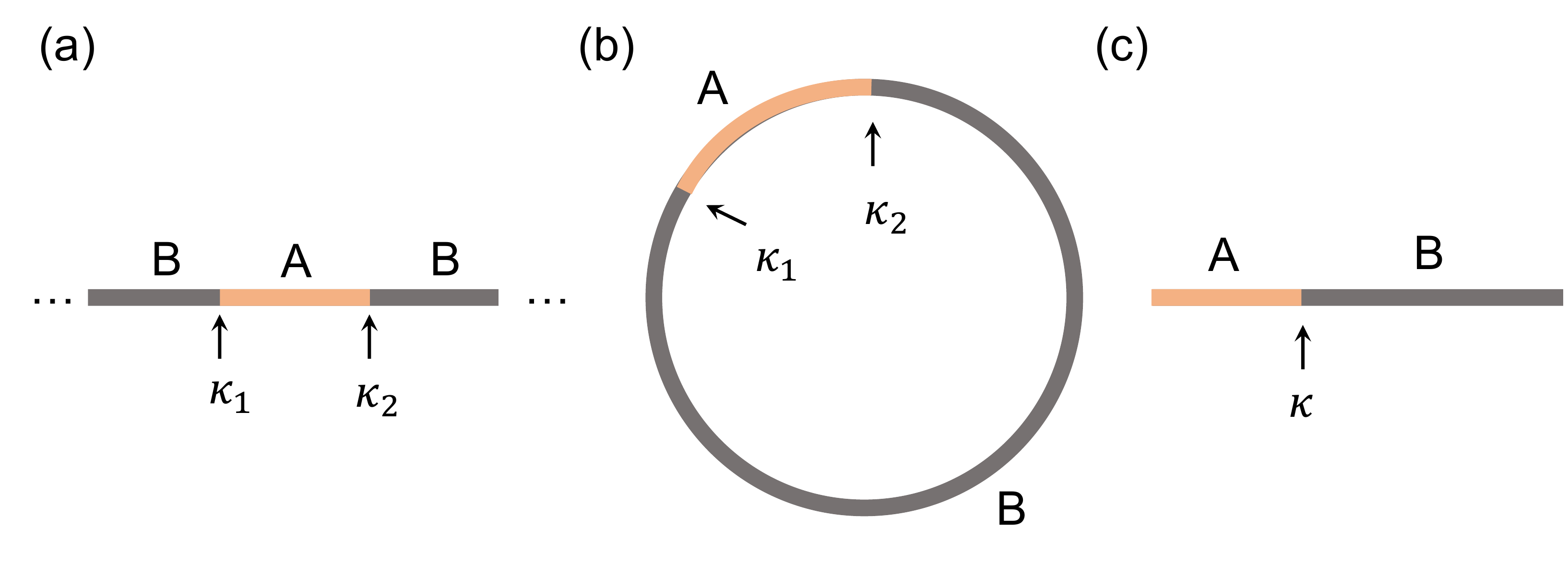}
\caption{\label{ent_geoms} Schematic of the three different geometries considered: (a) a finite region~(A) of length $r$ in an infinite system,~(b) a finite region of length $r$ in a periodic ring of length $L$ and~(c) a finite region of length~$r$ in a finite system of length $L$. In the last case, one end of the subsystem coincides with the end of the system. The cuts separating A from B are denoted by $\kappa_{1,2}$. Note that in panel (c), one of the entanglement cuts coincides with the physical termination of the total system, resulting in the geometry having only one entanglement cut. This leads to a different leading order dependence of the EE~[see Eqs.~(\ref{EE_per_ring},\ref{EE_open_chain})].}
\end{figure}

Consider the ground state,~$|\Psi\rangle$, of a many-body system described by a CFT at zero temperature. Then, the EE of a spatial region, A, is the  von-Neumann entropy:
\begin{align}
S_A = -{\rm Tr}_A\big(\rho_A\ln\rho_A\big) = -\lim_{n\to1}\frac{\partial}{\partial n}{\rm Tr}\rho_A^n,
\end{align}
Here $n$ is the replica index and $\rho_A = {\rm Tr}_B(|\Psi\rangle\langle \Psi|)$. Furthermore, B denotes the rest of the system. The density matrix $\rho_A$ can be computed using the path-integral formalism in euclidean space-time cut open at the intersection of the region A~(for a detailed derivation, see Ref.~\cite{Calabrese2009}). In particular, the EE can be obtained from a computation of the partition function $Z_1$ on the surface and $Z_n$ on its $n-$sheeted  cover~${\rm Tr}_A\rho_A^n = Z_n/Z_1^n$. For the geometries and models under consideration, this amounts to computation of the partition functions on annuli after suitable conformal transformations~\cite{Cardy2016}. We directly state the results for the EE for three different geometries~(see Fig.~\ref{ent_geoms}) and refer to the reader to Ref.~\cite{Cardy2016} for the derivations. 
In particular, for a finite region of length~$r$ within an infinite system~[Fig.~\ref{ent_geoms}(a)], the EE is given by
\begin{align}
\label{EE_inf_sys}
S_A = \frac{c}{3}\ln\frac{r}{a} + s_1 + s_2 + \cdots,
\end{align}
where $c$ is the central charge of the CFT and $a$ is the non-universal constant related to the UV cutoff. Furthermore, $s_{1,2} = \ln g_{1,2}$ are ${\cal O}(1)$ contributions to the EE that arise due to the `entanglement cuts',~$\kappa_{1,2}$, at the junction of the subsystem~(A) and the rest~(B). The $g_{1,2}$ are the corresponding $g$-functions~\cite{Affleck1991, Affleck1995}. For identical regularization procedures for the two cuts $\kappa_1$ and $\kappa_2$, $s_1 = s_2$. The dots correspond to subleading corrections~\cite{Cardy2016, Roy2020a}. Analogous results hold for a region within a periodic ring of length~$(L)$~[Fig.~\ref{ent_geoms}(b)], where the EE for subsystem A is given by:
\begin{align}
\label{EE_per_ring}
S_A = \frac{c}{3}\ln\Bigg[\frac{L}{\pi a}\sin\Big(\frac{\pi r}{L}\Big)\Bigg] + s_1 + s_2 + \ldots,
\end{align}
where the various variables are interpreted as before. Finally, we consider the case of a finite system of length $L$ with identical boundary conditions~\footnote{See Ref.~\cite{Cardy2016} for discussion on different boundary conditions at the two ends.} at the two ends, with one end of the subsystem coinciding with physical boundary of the total system~[Fig.~\ref{ent_geoms}(c)]. In this case, the EE is given by 
\begin{align}
\label{EE_open_chain}
S_A = \frac{c}{6}\ln\Bigg[\frac{2L}{\pi a}\sin\Big(\frac{\pi r}{L}\Big)\Bigg] + s_1 + s_2 + \ldots. 
\end{align}
Note that the coefficient of the leading logarithmic term is $c/6$ as opposed to $c/3$ of the previous cases. This accounts for the difference in the number of entanglement cuts in this case compared to the previous two. In particular, while the ${\cal O}(1)$-term $s_2$ arises from the entanglement cut $\kappa$, $s_1$ arises due to the physical boundary condition of the system. For generic systems, the boundary condition arising out of the entanglement cut is free boundary condition~\cite{Roy2020a, Cho2017, Roy2020b}. 

Below, we describe consider the two cases of the Ising and the free, compactified boson CFTs with different boundary conditions for the geometry in Fig.~\ref{ent_geoms}(c).

\subsection{Ising model}
\label{sec:Ising_BCFT}
\begin{figure}
\centering
\includegraphics[width = 0.4\textwidth]{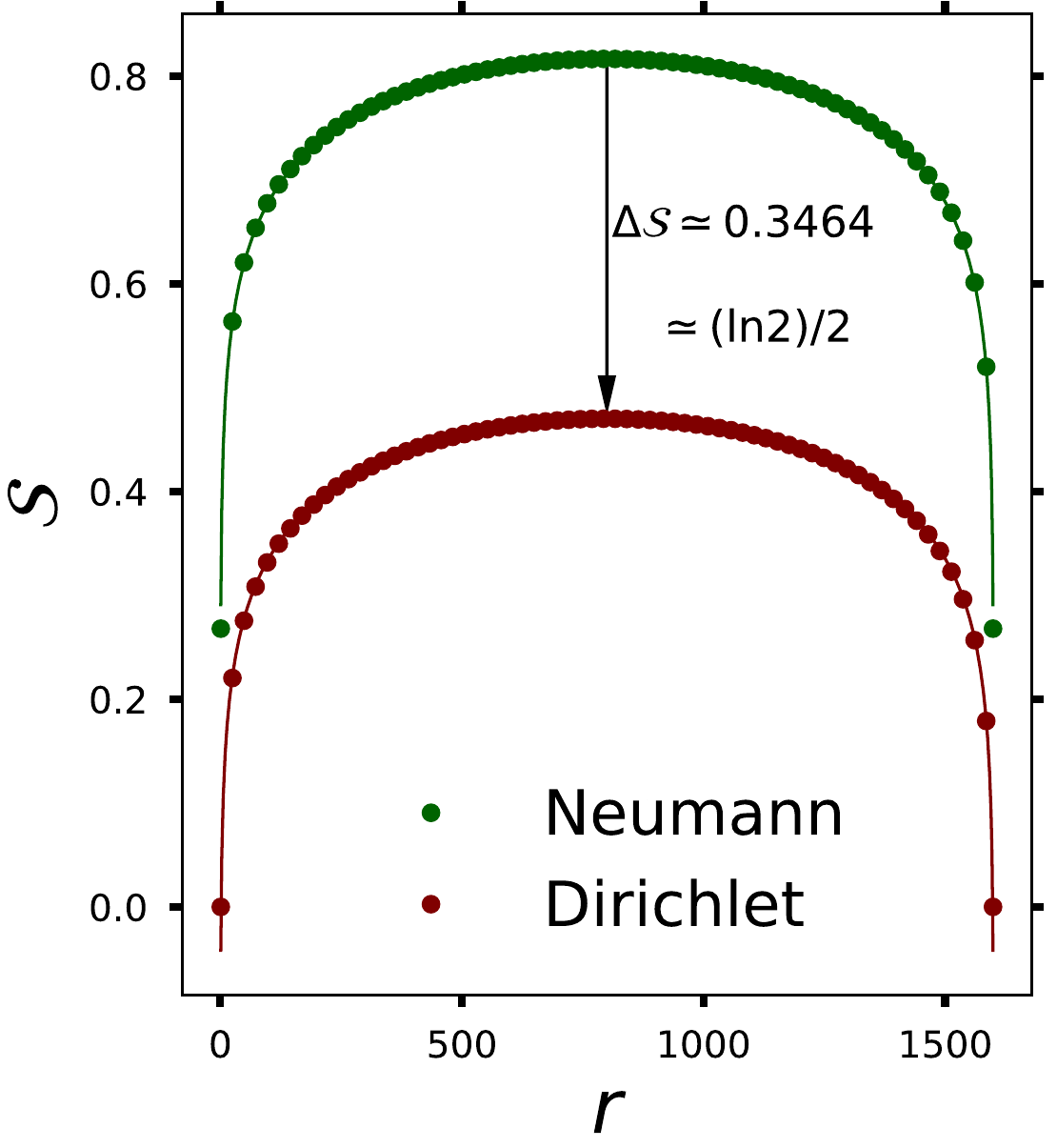}
\caption{\label{TFI_N_D} DMRG results for the critical transverse field Ising chain. The system size $L = 1600$. Entanglement entropy ${\cal S}$ as a function of the subsystem size $r$ for Neumann (green) and Dirichlet (maroon) boundary conditions. The central charge was verified to be $\simeq1/2$ by fitting the data for Neumann boundary conditions to Eq.~(\ref{EE_open_chain}). As the boundary condition changes from Neumann to Dirichlet, the entanglement entropy changes by $0.3464$ which is close to the expected change in the boundary entropy given by $(\ln 2)/2$.}
\end{figure}

The Ising model is the unitary, minimal model ${\cal M}(4,3)$ with central charge $c = 1/2$ (see, for example, Chapters 7 and 8 of Ref.~\cite{diFrancesco1997}). It contains three primary fields, $I, \sigma, \epsilon$, with conformal dimensions: $h_{I}=0, h_{\sigma}=1/16$ and $h_{\epsilon}=1/2$.
We will consider two cases: (i) free/Neumann (N) boundary conditions at both ends and (ii) fixed/Dirichlet (D) boundary conditions at both ends. The boundary states for the different boundary conditions are given by~\cite{Cardy1989}
\begin{align}
|\tilde{0}\rangle &= \frac{1}{\sqrt{2}}|0\rangle + \frac{1}{\sqrt{2}}|\epsilon\rangle + \frac{1}{2^{1/4}}|\sigma\rangle,\nonumber\\
\Big|\tilde{\frac{1}{2}}\Big\rangle &= \frac{1}{\sqrt{2}}|0\rangle + \frac{1}{\sqrt{2}}|\epsilon\rangle - \frac{1}{2^{1/4}}|\sigma\rangle,\nonumber\\\label{bs_Ising}
\Big|\tilde{\frac{1}{16}}\Big\rangle &= |0\rangle - |\epsilon\rangle,
\end{align}
where the first two correspond to Dirichlet boundary conditions and the last corresponds to Neumann boundary condition~(see Chap. 11 of Ref.~\cite{diFrancesco1997} for further details). The corresponding $g-$functions are 
\begin{align}
g_N = \Big\langle 0\Big|\tilde{\frac{1}{16}}\Big\rangle = 1,\ g_D =  \langle0|\tilde{0}\rangle = \Big\langle 0\Big|\tilde{\frac{1}{2}}\Big\rangle=\frac{1}{\sqrt{2}}. 
\end{align}
This straightforwardly reveals the the change in boundary entropy as the boundary condition is changed from Neumann to Dirichlet:
\begin{align}
\label{delta_S}\Delta S = s_N - s_D = \ln\frac{g_N}{g_D}=\frac{1}{2}\ln2. 
\end{align}
Next, we compute using DMRG, the EE in a critical transverse-field Ising chain and compare with the analytical CFT predictions derived above. The Neumann condition is implemented by a spin chain with open boundaries. The lattice Hamiltonian is given by 
\begin{align}
\label{H_TFI_N}
H_{\rm TFI}^{N} = -\frac{1}{2}\sum_{i=1}^{L-1}\sigma_i^x\sigma_{i+1}^x - \frac{1}{2}\sum_{i=1}^L\sigma_i^z. 
\end{align}
The Dirichlet boundary condition is implemented by adding a small longitudinal boundary field at the two ends of the chain:
\begin{align}
\label{H_TFI_D}
H_{\rm TFI}^D = H_{\rm TFI}^{N} - h_b(\sigma_1^x + \sigma_L^x).
\end{align}
The boundary field sets a correlation length-scale. By looking at the change in the EE for a subsystem size much larger than this correlation length, we can extract the boundary entropy contribution in the scaling limit. 
In our simulations, we chose the system size to be $L=1600$ and a bond-dimension of $600$ to keep truncation errors below $10^{-12}$. We verify the central charge ($c$) to be $\simeq1/2$. This is done by evaluating the entanglement entropy ${\cal S}$ for a finite block (of length $r$) within the system (of length $L$) and fitting to Eq.~\eqref{EE_open_chain} for Neumann and Dirichlet boundary conditions at both ends of the chain. 
By changing the boundary conditions, we obtain a change in entropy that is very close to the expected value of $(\ln2)/2$~(see Fig.~\ref{TFI_N_D}). 
 
\subsection{The free, compactified boson model}
\label{sec:free_boson_BCFT}
Consider the free, compactified boson CFT with a compactification radius~$R$ over an interval $[0,L]$ with either Neumann or Dirichlet boundary conditions. The euclidean action is given by
\begin{align}\label{free_bos_action}
{\cal A}_0 = \frac{1}{2}\int_0^\beta dt\int_0^L dx\Big[(\partial_t\phi)^2 + (\partial_t\phi)^2\Big],
\end{align}
where $\beta$ is the inverse temperature. At the boundary, Neumann boundary condition corresponds to $\partial_x\phi = 0$, while Dirichlet corresponds $\phi=\phi_0$, where $\phi_0$ is a constant. The boundary states for this model are well-known~\cite{Callan:1987px, Oshikawa1997}. They are 
\begin{align}
|N(\tilde{\phi}_0)\rangle &= \sqrt{R\sqrt{\pi}}\sum_{m}e^{-\frac{im\tilde{\phi}_0R}{2\sqrt{\pi}}}{\rm{exp}}\Big[+\sum_{k>0}\tilde{a}_{-k}^\dagger \tilde{a}_{k}^\dagger\Big]|0,m\rangle,\\
|D(\phi_0)\rangle &= \frac{1}{\sqrt{2R\sqrt{\pi}}}\sum_{n}e^{-\frac{in\phi_0}{R\sqrt{\pi}}}{\rm{exp}}\Big[-\sum_{k>0}\tilde{a}_{-k}^\dagger \tilde{a}_{k}^\dagger\Big]|n,0\rangle.
\end{align}
Note the duality between Neumann and Dirichlet boundary conditions: $R\leftrightarrow2/R$ and $\tilde{\phi}_0$ is the field dual to $\phi_0$. Recall that the integers $m, n$ determine the winding number and the quantization of the zero-mode momenta respectively. The normalizations for the boundary states directly yield the corresponding $g$-functions leading to the following change in the boundary entropies: 
\begin{align}
g_N = \sqrt{R\sqrt{\pi}}, g_D = \frac{1}{\sqrt{2R\sqrt{\pi}}}\Rightarrow \Delta S = \ln\frac{g_N}{g_D} = \frac{1}{2}\ln(2R^2\pi).
\end{align}

Next, we provide numerical verification of the above result. To that end, we consider a finite XXZ spin chain. The open spin chain realizes the Neumann boundary condition:
\begin{align}
\label{H_XXZ_N}
H_{\rm XXZ}^N = -\frac{1}{2}\sum_{i=1}^{L-1}\Big[\sigma_i^x\sigma_{i+1}^x + \sigma_i^y\sigma_{i+1}^y + \Delta \sigma_i^z\sigma_{i+1}^z\Big],
\end{align}
where $\Delta$ is the anisotropy parameter. As is well-known, the long-wavelength properties of this spin chain, in the paramagnetic regime, $-1\leq \Delta\leq1$, is well-described by the free, compactified boson CFT~[see Eq.~\eqref{free_bos_action}]~\cite{Lukyanov1997, Lukyanov1999}. The compactification radius is related to the Luttinger parameter~($K$) in the following way: 
\begin{align}
\label{K_def}
R = \frac{1}{\sqrt{\pi K}},\ K = \frac{2}{\pi}\cos^{-1}\Delta.
\end{align}
The Dirichlet boundary condition is realized by adding a small-transverse field along the $\sigma_x$ direction at the boundaries:
\begin{align}
\label{H_XXZ_D}
H_{\rm XXZ}^D = H_{\rm XXZ}^N - h_b(\sigma_1^x + \sigma_{L}^x),
\end{align}
where $h_b$ is the boundary field strength. Bosonizing the Hamiltonian with boundary fields leads to the boundary sine-Gordon Hamiltonian~\cite{Ghoshal1994, Fendley1994} with massless bulk~\cite{Caux:2003mw} in the scaling limit~(we follow the bosonization rules of Ref.~\cite{Lukyanov2003}):
\begin{align}
{\cal A}_{\rm bSG} = {\cal A}_0 + M_b\int dt \Bigg[\cos\frac{\beta\phi(x=0)}{2} + \cos\frac{\beta\phi(x=L)}{2}\Bigg],
\end{align}
where $\beta = \sqrt{4\pi K}$ and $M_b$ is the boundary potential strength which depends on the lattice parameter $h_b$. A nonzero $M_b$~(equivalently, a nonzero $h_b$) induces the boundary RG flow from Neumann to Dirichlet boundary conditions. In terms of the Luttinger parameter, the corresponding change in the boundary entropy is given by
\begin{align}
\label{DS_N_D}
\Delta S = \frac{1}{2}\ln\frac{2}{K}. 
\end{align}
\begin{figure}
\centering
\includegraphics[width = \textwidth]{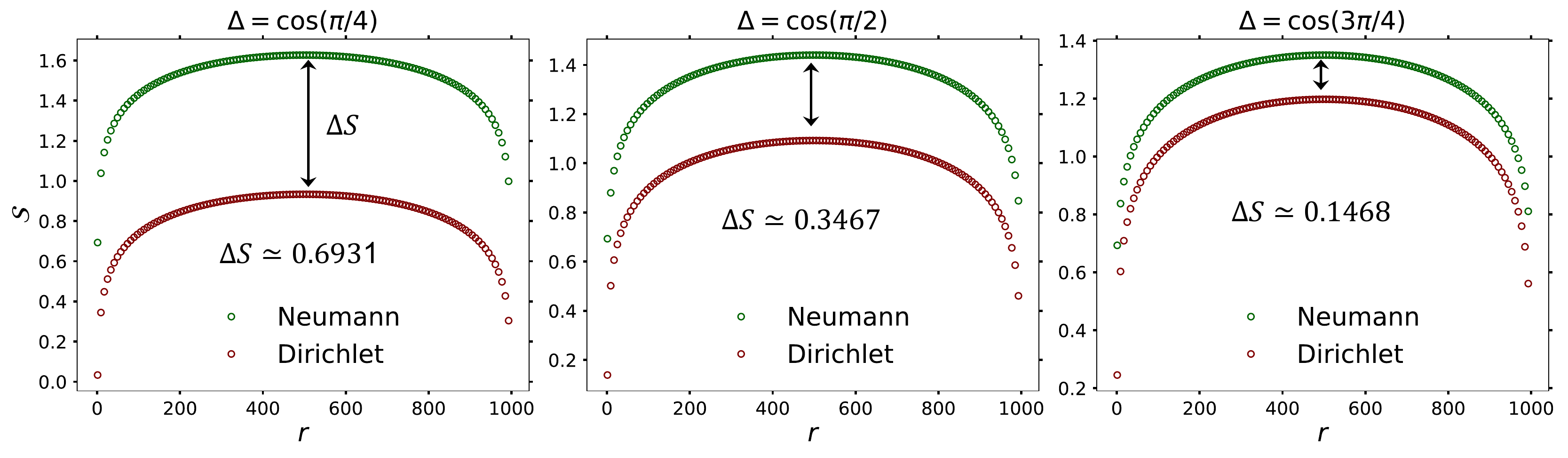}
\caption{\label{XXZ_N_D} DMRG results for the XXZ chain. The anisotropy parameters are $\Delta = \cos(\pi/4)$~(left), $\cos(\pi/2)$~(center) and $\cos(3\pi/4)$~(right). The corresponding Luttinger parameters are $K = 0.5, 1.0$ and $K = 1.5$. The system size $L = 1000$. Entanglement entropy, $S$, as a function of the subsystem size $r$ for Neumann (green) and Dirichlet (maroon) boundary conditions. The central charge was verified to be $\simeq1$ by fitting to Eq.~(\ref{EE_open_chain}). As the boundary condition changes from Neumann to Dirichlet, the EE changes by $0.6931$, $0.3467$ and $0.1468$ respectively. The obtained results are close to the expected changes in the boundary entropy: $\ln2, (\ln 2)/2$ and $[\ln(4/3)]/2$ respectively.}
\end{figure}
Fig.~\ref{XXZ_N_D} shows the results of the computation of the EE for the various bipartitionings of the system for $\Delta = \cos(\pi/4), \cos(\pi/2)$ and $\cos(3\pi/4)$. The corresponding Luttinger parameters are given by $K = 1/2, 1$ and $3/2$. The change in the boundary entropy is computed by taking the difference of the EE values near the center of the chain. The obtained~[expected] values of $\Delta S$ are $\simeq0.6931~[\ln2]$, $0.3467~[(\ln2)/2]$ and $0.1468~[\{\ln(4/3)\}/2]$ for the chosen values of the $\Delta$. 

\section{Entanglement Entropy in CFTs with defects}
\label{sec:EE_dCFT}
In this section, we consider the more general setting of two CFTs glued together by a defect and investigate the behavior of the interface EE, {\it i.e.}, the EE across the defect. The latter measures the amount of entanglement between the two CFTs glued together at the defect as long as the state of the total system remains pure. Intuitively, the presence of the defect leads to back-scattering of the information-carrying modes of the system. This leads to a diminishing of entanglement between the left and right halves of the system connected at the defect compared to the case when there is no defect. In particular, we concentrate on those defects which are marginal perturbations around fixed point without any defects. For these models, the interface EE exhibits still the logarithmic scaling characteristic of the CFT without defects. However, unlike the case without defects, the coefficient of the logarithmic scaling yields a continuously-varying `effective central charge' -- a manifestation of the well-known effect that marginal perturbations lead to continuously-varying scaling exponents~\cite{Cardy1987, Igloi1993}. In fact, the central charge depends continuously on the transmission coefficient,~$t$, of the scattering matrix in the scattering picture. 

Next, we describe the behavior of the interface EE for the Ising and the free, compactified boson models in the presence of such defects. We will always consider the case where the defect lies at the center of a chain with open boundary conditions and compute the interface EE for a subsystem extending from the left end of the system up to the defect. Then, the interface EE~($S_{\cal I}$) for a total system-size $L$ scales as:
\begin{align}
\label{iEE_sc}
S_{\cal I}[t(b)] = \frac{c_{\rm eff}[t(b)]}{6}\ln \frac{L}{a} + s_1 + s_2[t(b)] + \cdots,
\end{align}
where $c_{\rm eff}[t(b)]$ is a continuously-varying `effective central charge' and $b$ is the defect strength. It occurs with a factor $1/6$ since there is effectively only one entanglement cut for a system with open boundary conditions~(see discussion in Sec.~\ref{sec:EE_BCFT}). The subleading term has two contributions: $s_1$ arises from the boundary condition on the left and $s_2[t(b)]$ from the defect. Finally, the dots indicate terms that are smaller than ${\cal O}(1)$. The explicit dependence of the $c_{\rm eff}$ on the defect strength is nontrivial and has been analytically obtained for the free, real fermion~\cite{Eisler2010, Brehm2015} and the free, compactified boson~\cite{Sakai2008, Peschel2012e}. They are provided below. 

\subsection{The Ising model}
\label{sec:Ising_dCFT}
In this section, we describe the two defect classes: energy and duality of the Ising model. 
\subsubsection{Energy defect}
\label{sec:Ising_dCFT_e}
The energy defect for the Ising model arises due to a ferromagnetic coupling with an altered strength~$b_\epsilon$~\cite{Oshikawa1997, Kadanoff1971}. The Hamiltonian is given by 
\begin{align}
\label{TFI_def_e}
H_{\rm TFI}^\epsilon = -\frac{1}{2}\sum_{j=1}^{L-1}\sigma_j^x\sigma_{j+1}^x -\frac{1}{2}\sum_{j=1}^L\sigma_j^z + \frac{1-b_\epsilon}{2}\sigma_{i_0}^x\sigma_{i_0+1}^x.
\end{align}
where $i_0 = L/2$~(for definiteness, we take $L$ even). 
For our purposes, it is sufficient to consider $b_\epsilon\in[-1,1]$. In particular, $b_\epsilon=1$ corresponds to the case when there is no defect, while $b_\epsilon = 0$ splits the chain in two halves~\footnote{Note that the open chain can be obtained from the periodic Ising chain by introducing an energy defect between the sites $L$ and 1 with strength $b_\epsilon = 0$.}. Finally, $b_\epsilon = -1$ corresponds to an antiferromagnetic bond in the middle of the chain. Note that for an open chain, unlike for a periodic chain, any defect Hamiltonian with defect strength $-b_\epsilon$ can be transformed to one with defect strength $b_\epsilon$ under a nonlocal unitary transformation. The latter involves flipping all the spins on one half of the chain. In this way, a~$b_\epsilon=-1$ defect can be transformed away to the case without a defect. 

As is evident from Eq.~\eqref{TFI_def_e}, the defect part of the Hamiltonian is indeed a marginal perturbation by the primary operator~$\epsilon$ at a point in space~(recall that the conformal dimensions of $\epsilon$ are $h_\epsilon=\bar{h}_\epsilon=1/2$). This model can be mapped, using standard folding maneuvers~\cite{Oshikawa1997, Saleur1998}, to a boundary problem of the $\mathbb{Z}_2$ orbifold of the free-boson. This allows computation of relevant spin-spin correlation functions across the defects~\cite{Oshikawa1997}. Since our interest is in the interface EE, we use a different, exact, solution of the problem by mapping it to a fermionic model. The Hamiltonian of the latter model is bilinear in fermionic creation and annihilation operators and can be diagonalized semi-analytically~\cite{Henkel1989, Baake1989}. This leads to very efficient computation of EE from the ground-state correlation matrix using techniques of Refs.~\cite{Vidal2002,Peschel2003, Latorre2004}. 

\begin{figure}
\centering
\includegraphics[width = 0.9\textwidth]{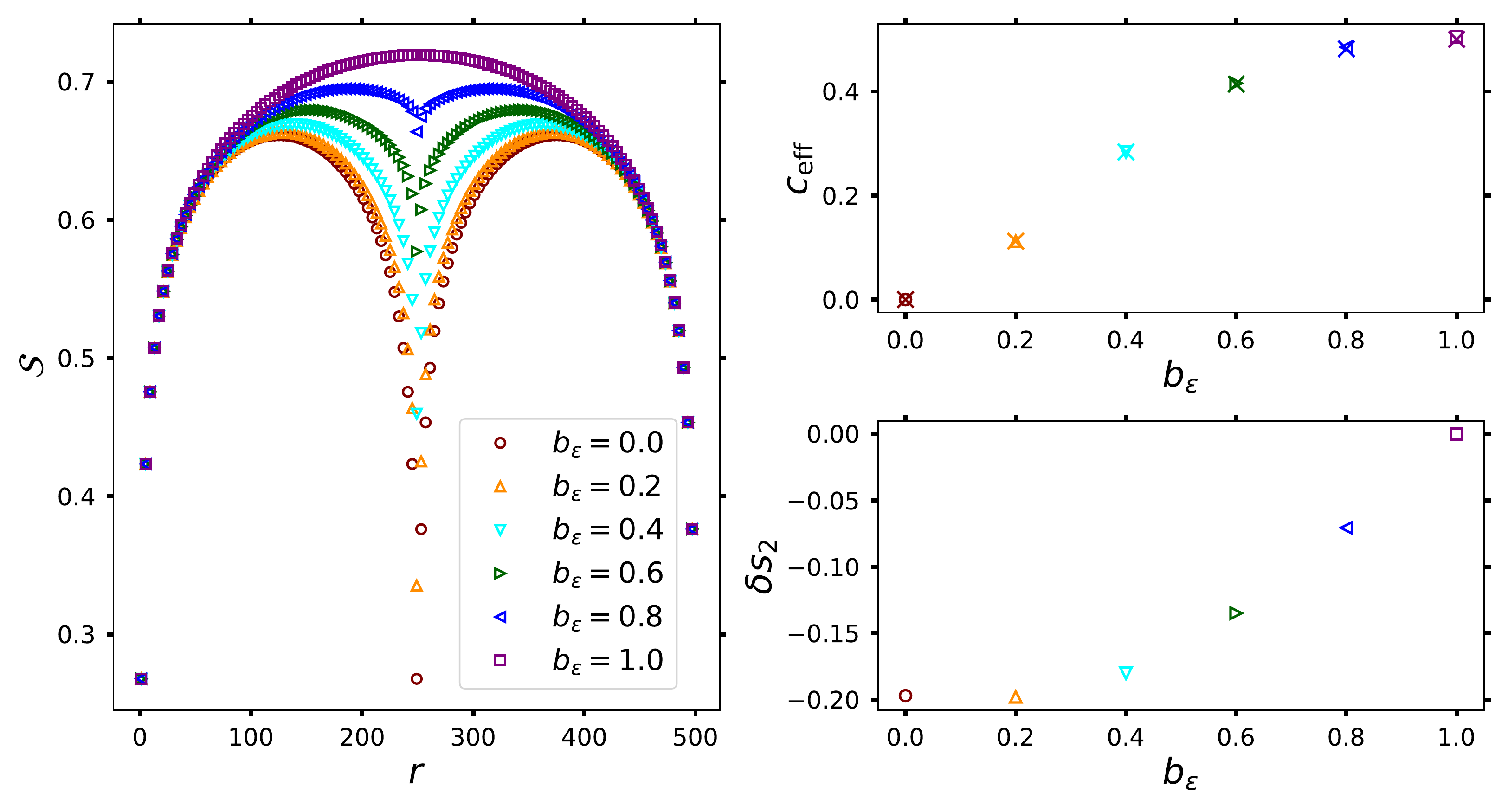}
\caption{\label{TFI_inter_entropy_e} Results for EE~($S$) for an open Ising chain with an energy defect. The defect strength is varied from 0 to 1 in steps of 0.2.~(Left) The EE for different bipartitionings of the in a system, with a total system size,~$L=500$. The dip around the center of the chain for the EE is due to the defect.~(Top right) Effective central charge from the scaling of the interface EE~($S_{\cal I}$) for varying system sizes from $L =100$ to $L = 500$. Fitting to Eq.~\eqref{iEE_sc} yields the corresponding $c_{\rm eff}$-s, which are plotted in the top right panel. For comparison, the analytical predictions from Eq.~\eqref{ceff} are shown with crosses of the corresponding color. The corresponding offsets, normalized with respect to the case of no defect:~$\delta s_2 = s_2[t(b_\epsilon)] - s_2[t(b_\epsilon = 1)]$ are obtained from the linear fit and are plotted in the bottom right panel. We do not know of an analytical expression for the $\delta s_2$-s. }
\end{figure}

This is done using the Jordan-Wigner~(JW) transformation: 
\begin{align}
\label{jw}
\gamma_{2k-1} = \sigma_k^x\prod_{j=1}^{k-1}\sigma_j, \ \gamma_{2k} = \sigma_k^y\prod_{j=1}^{k-1}\sigma_j,
\end{align}
where $\gamma_{j}$-s are real, fermion operators obeying $\{\gamma_j, \gamma_k\} = 2\delta_{j,k}$. Note that $-i\gamma_{2k-1}\gamma_{2k}=\sigma_k^z$. The resulting fermionic Hamiltonian is given by
\begin{align}
\label{TFI_def_e_f}
H_{\rm TFI}^{\epsilon, f} =  \frac{i}{2}\sum_{j=1}^{L-1}\gamma_{2j}\gamma_{2j+1} + \frac{i}{2}\sum_{j=1}^L\gamma_{2j-1}\gamma_{2j} -\frac{i(1-b_\epsilon)}{2}\gamma_{2i_0}\gamma_{2i_0 + 1}. 
\end{align}
Now, we diagonalize this Hamiltonian numerically and compute the EE for different bipartionings of the system by computing the ground-state correlation matrix~(similar results have been obtained in Ref.~\cite{Igloi2009}). Fig.~\ref{TFI_inter_entropy_e}(left) shows the EE for varying $b_\epsilon$ from 0 to 1.0 in steps of 0.2. The interface EE is obtained by choosing $r = L/2$. The scaling of the interface EE with $L$ for $b_\epsilon = 0.2$ is shown on the top right panel. Fitting to Eq.~\eqref{iEE_sc} leads to a $c_{\rm eff} = 0.112$. The analytical prediction for the effective central charge is~\cite{Eisler2010, Brehm2015} 
\begin{align}
\label{ceff}
c_{\rm eff}(t) &= \frac{|t|}{2} - \frac{1}{2} - \frac{3}{\pi^2}\Big[(|t|+1)\ln(|t|+1)\ln(|t|) + (|t|-1){\rm Li}_2(1-|t|) + (|t|+1){\rm Li}_2(-|t|)\Big],
\end{align}
where $t = \sin[2(\cot^{-1}b_\epsilon)]$ and ${\rm Li}_2$ is the dilogarithm function~\cite{Lewin1981}. The results obtained by the free-fermion technique is compatible with that predicted by Eq.~\eqref{ceff} up to the third decimal place. Fig.~\ref{TFI_inter_entropy_e} shows a comparison of the effective central charges~(top right panel) and the corresponding offsets~(bottom right panel). We compute the offsets normalized with respect to the case of no-defect and plot $s_2[t(b_\epsilon)] - s_2[t(b_\epsilon = 1)]$ for different values of $b_\epsilon$. We do not know of an analytical expression for the normalized offsets. 
\subsubsection{Duality defect}
\label{sec:Ising_dCFT_d}
The duality defect~\cite{Aasen2016, Grimm2001} of the Ising model between the sites at $i_0$ and $i_0 + 1$ arises due to an interaction of the form $\sigma_{i_0}^x\sigma_{i_0 + 1}^y$ instead of the usual ferromagnetic coupling. Equally important, there is no transverse field at the site $i_0 + 1$. The resulting defect Hamiltonian is given by
\begin{align}
\label{TFI_def_d}
H_{\rm TFI}^\sigma = -\frac{1}{2}\sum_{\substack{j=1,\\ j\neq i_0}}^{L-1}\sigma_j^x\sigma_{j+1}^x -\frac{1}{2}\sum_{\substack{j=1,\\ j\neq i_0 + 1}}^L\sigma_j^z -\frac{b_\sigma}{2}\sigma_{i_0}^x\sigma_{i_0+1}^y.
\end{align}
The duality defect for $b_\sigma = 1$~(equivalently $b_\sigma = -1$, which is related by a local unitary rotation) is the topological defect for the Ising CFT. Note that this duality defect Hamiltonian is related by a local unitary rotation on the $(i_0+1)^{\rm th}$ spin to the one considered in Refs.~\cite{Oshikawa1997}, which has $\sigma_i^x\sigma_{i_0+1}^z$ interaction. We do not use this alternate form since it no longer leads to a bilinear Hamiltonian under JW transformation and cannot be solved by the free-fermion technique. 

The fundamental difference between the duality and the energy defect Hamiltonians is best captured by the JW transformation. In the fermionic language, the defect Hamiltonian is given by
\begin{align}
\label{TFI_def_e_f}
H_{\rm TFI}^{\sigma, f} &=  \frac{i}{2}\sum_{\substack{j=1,\\ j\neq i_0}}^{L-1}\gamma_{2j}\gamma_{2j+1} + \frac{i}{2}\sum_{\substack{j=1,\\ j\neq i_0 + 1}}^{L-1}\gamma_{2j-1}\gamma_{2j} + \frac{ib_\sigma}{2}\gamma_{2i_0}\gamma_{2i_0+2}.
\end{align}
Note that the operator $\gamma_{2i_0+1}$ does not occur in $H_{\rm TFI}^{\sigma, f}$. It commutes with the Hamiltonian: $[\gamma_{2i_0+1}, H_{\rm TFI}^{\sigma,f}] = 0$, and anticommutes with the conserved $\mathbb{Z}_2$ charge: $\{\gamma_{2i_0+1},Q\} =0$, where $Q=\prod_{j=1}^L\sigma_j^z = 1$. Thus, it is a zero-mode of the model which is perfectly localized in space. It has a partner zero-mode which is completely delocalized: 
\begin{align}
\Lambda(b_\sigma) = b_\sigma\sum_{k-1}^{i_0}\gamma_{2k-1} + \sum_{k=i_0+1}^L\gamma_{2k}.
\end{align}
Note that the zero-modes exist for all values of $b_\sigma$ and are not special features of the topological point. The fermionic Hamiltonian also reaffirms a CFT result~\cite{Grimm1992, Grimm2001}:~$H_{\rm TFI}^{\sigma, f}$ describes a chain of $2L-1$ Majorana fermions or equivalently, $L-1/2$ spins. This is important for quantifying finite-size effects. 
\begin{figure}
\centering
\includegraphics[width = 0.9\textwidth]{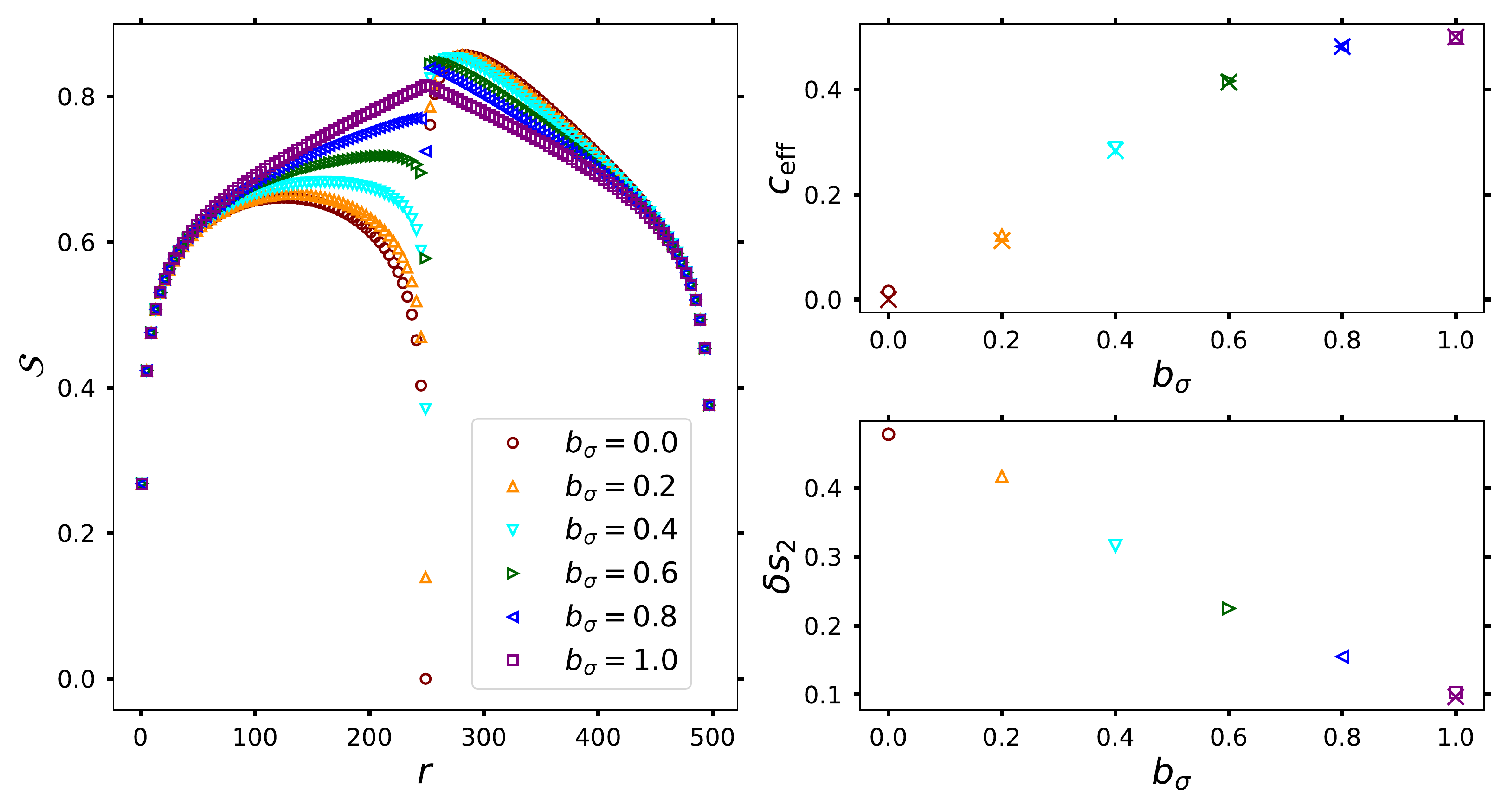}
\caption{\label{TFI_inter_entropy_d} Results for EE~($S$) for an open Ising chain with a duality defect. The defect strength is varied from 0 to 1 in steps of 0.2.~(Left) The EE for different bipartitionings of the in a system, with a total system size,~$L=500$. The dip around the center of the chain for the EE is due to the defect.~(Top right) Effective central charge from the scaling of the interface EE~($S_{\cal I}$) for varying system sizes from $L =100$ to $L = 500$. Fitting to Eq.~\eqref{iEE_sc} yields $c_{\rm eff}$, which are plotted in the top right panel. For comparison, the analytical predictions from Eq.~\eqref{ceff} are shown with crosses of the corresponding color. The difference in offsets~[$\delta s_2 = s_2(b_\sigma) - s_2(b_\epsilon = 1)$] from the linear fit are plotted in the bottom right panel. Only for $b_\sigma = 1$, the $\delta s_2$ is known exactly and is given by $\Delta S(1/2)/2= -1/4 + (\ln2)/2$~\cite{Roy2021a}. }
\end{figure}

Now, we compute the EE for different bipartitionings of the system from the ground state correlation matrix. However, unlike the energy defect for an open chain, as was just described, there are zero-energy states of the defect Hamiltonian. The existence of these states raises the question: are the zero-energy states empty or occupied in the ground state? Yet another possibility is to consider an incoherent superposition of filled and empty states. The latter possibility leads to the total system being in a mixed state, but is appropriate when taking the zero-temperature limit of a thermal ensemble~\cite{Herzog2013}. The question is crucial to the computation since zero-energy modes nontrivially affect the EE. In fact, for a periodic chain of free, real, fermions, when the total system is in a mixed state, the zero-modes give rise to nontrivial corrections to the EE of a subsystem of size~$r$ within a total system of size~$L$~\cite{Herzog2013, Klich2017}. The correction is given by
\begin{align}
\Delta S \Big(\frac{r}{L}\Big) =\frac{\pi r}{L} \int_0^\infty \tanh\Big(\frac{\pi r h}{L}\Big)[\coth(\pi h) - 1]. 
\end{align}
For $r\ll L$, the EE is oblivous to the existence of the two nonlocal zero-modes spread throughout the system:~$\Delta S\simeq \pi^2r^2/12L^2\rightarrow 0$. The situation changes as the subsystem occupies appreciable fraction of the total system~($r\sim L$) culminating in $\Delta S(r = L) = \ln2$, the latter being the entropy of the two-fold degenerate ground state of a periodic chain of free, real fermions. Similar corrections occur for the Ising chain with the duality defect for different choices of the total system being pure or mixed~\cite{Roy2021a}. Below, we describe the results for the case when the total system is in a pure state and refer the reader to Ref.~\cite{Roy2021a} for the mixed state results. 

Fig.~\ref{TFI_inter_entropy_d}(left) shows the results for the EE for various bipartitionings of the system of size 500. The strength of the defect~($b_\sigma$) is varied from 0.0 to 1.0 in steps of 0.2. The scaling of the interface EE with different system sizes $L$ yields, as for the energy defect, the effective central charge~[see Eq.~\eqref{iEE_sc}]. The obtained values of $c_{\rm eff}$ are shown in the top right panel, with the expected values from Eq.~\eqref{ceff}. The offsets from this fit, normalized with respect to the case~$b_\epsilon=1$, are shown in the bottom right panel. Analytical result for the offset is known only for $b_\sigma=1$, when $\delta s_2 = \Delta S(1/2)/2 = -1/4 + (\ln2)/2$~\cite{Roy2021a}. 

\subsection{The free, compactified boson model}
\label{sec:free_boson_dCFT}
In this section, we describe conformal interfaces of two free, compactified boson CFTs with different compactification radii $R_A$ and $R_B$~\cite{Affleck2001, Bachas2001}. Unlike the Ising case, the `defect' is extended throughout one half of the system~(see below for a lattice realization). For $R_A\neq R_B$, the EE across the interface is lower compared to the case $R_A=R_B$. This can be again understood due to the reflections of the incident wave at the interface for $R_A\neq R_B$. The relevant quantity is the scattering matrix which can be derived by a variety of methods~\cite{Bachas2001}. Below, we present an intuitive explanation based on elementary notions of electrical engineering. 

The free, compact boson model with compactification radius~$R_\alpha$ can be viewed as describing a quantum transmission line~\cite{Goldstein2013, Roy2019, Roy2020a, Roy2020b}, where the impedance~($Z_\alpha$) of the $\alpha^{\rm th}$ line is related to the compactification radius as: $Z_\alpha\sim 1/R_\alpha^2$. At the interface of two transmission lines with impedances $Z_A, Z_B$, the reflection coefficient for incoming waves is given by~\cite{Pozar1990, Clerk2010}
\begin{align}
r = \frac{Z_B-Z_A}{Z_B + Z_A} = \frac{R_A^2-R_B^2}{R_A^2+R_B^2} = \cos(2\theta),\ \theta = \tan^{-1}\frac{R_B}{R_A}.
\end{align}
The corresponding transmission coefficient is given by $t = \sin(2\theta)$ with $|r|^2+|t|^2=1$. As for the Ising model, the interface EE is determined by the transmission coefficient $t$~[see Eq.~\eqref{iEE_sc}]. However, the explicit form of the central charge is different from the Ising case and is given by~\cite{Sakai2008, Peschel2012e}:
\begin{align}
\label{ceff_bos}
c_{\rm eff}(|t|) &= \frac{1}{2} + |t| + \frac{3}{\pi^2}\Big[(|t|+1)\ln(|t|+1)\ln |t| + (|t|-1){\rm Li}_2(1-|t|) + (|t|+1){\rm Li}_2(-|t|)\Big].
\end{align} 

This conformal interface is realized on the lattice by two XXZ chains with anisotropies $\Delta_A$ and $\Delta_B$ on the two sides of the interface. Thus, the defect Hamiltonian is given by
\begin{align}
\label{H_XXZ_int}
H_{\rm XXZ}^{\cal I} &= -\frac{1}{2}\sum_{i=1}^{L-1}\Big[\sigma_i^x\sigma_{i+1}^x + \sigma_i^y\sigma_{i+1}^y + \Delta_i \sigma_i^z\sigma_{i+1}^z\Big], \ \Delta_i = \Delta_A\theta(i_0-i) + \Delta_B\theta(i-i_0),
\end{align}
where $\theta$ is the Heaviside theta function. The role of the impedance is played the Luttinger parameter times the impedance quantum~($6.5k\Omega$). Thus, 
\begin{align}
r = \frac{K_B - K_A}{K_B + K_A},\ \theta = \tan^{-1}\sqrt{\frac{K_A}{K_B}}.
\end{align}
As expected, for $K_A=K_B$, which corresponds to $\Delta_A = \Delta_B$, the interface disappears with $r = 0, t = 1$ and $c_{\rm eff}(1) = 1$. On the other hand, for $\Delta_A\neq\Delta_B$, either $\Delta_A = 1$ or $\Delta_B= 1$ corresponds to $r = 1, t = 0$ and $c_{\rm eff}(0) = 0$. 
 \begin{figure}
\centering
\includegraphics[width = 0.9\textwidth]{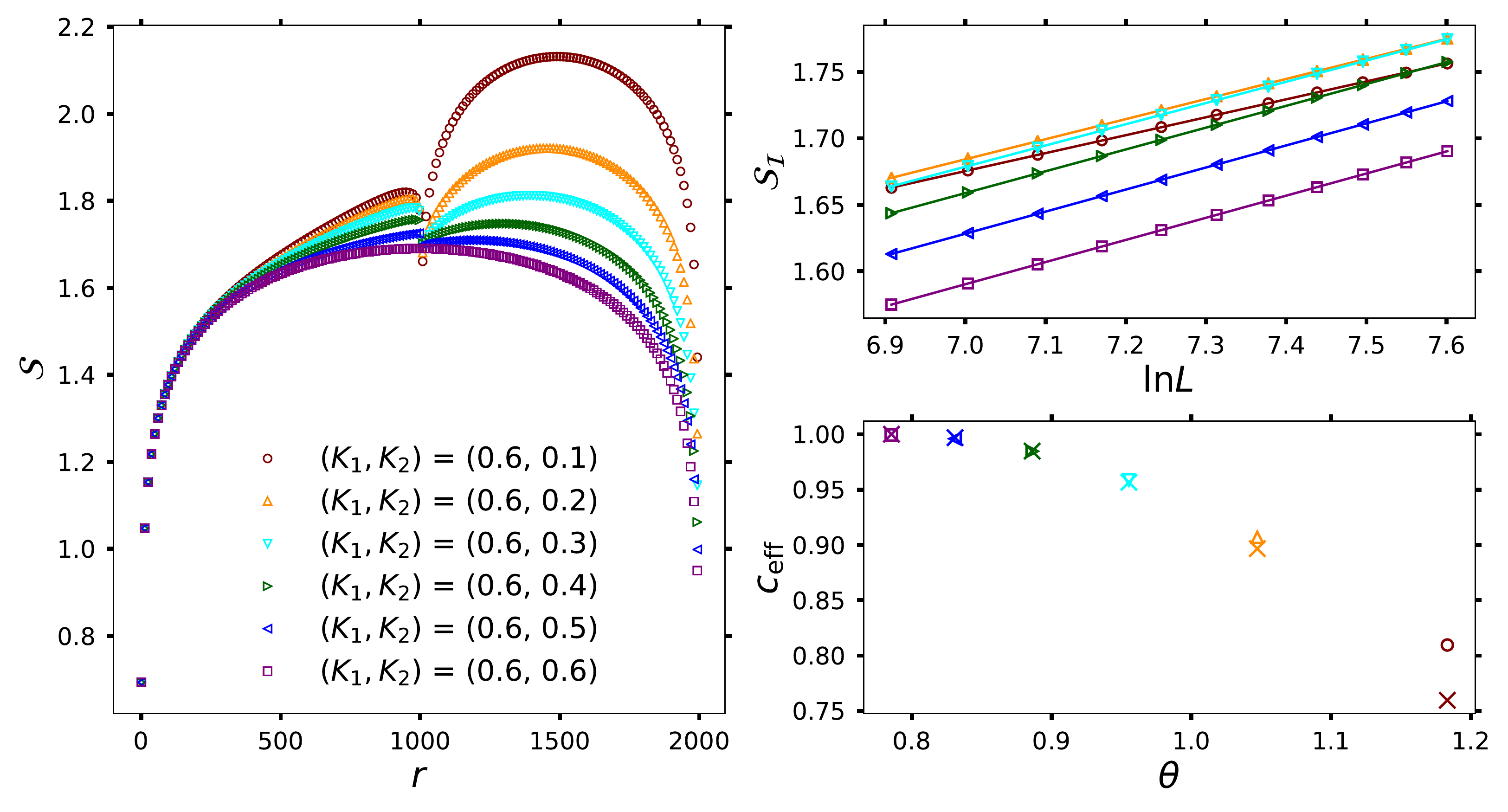}
\caption{\label{XXZ_inter_entropy} Results for interface EE~($S$) for an open XXZ chain with an interface defect. The left half of the chain has a Luttinger parameter~$K_1 = 0.6$, while the same for the right half~($K_2$) is varied from 0.1 to 0.6~[see Eq.~\eqref{K_def} for relation to the anisotropy parameters].~(Left) The EE for different bipartitionings of the in a system, with a total system size,~$L=2000$. The dip around the center of the chain for the EE is due to the defect.~(Top right) Scaling of the interface EE~(${\cal S}_{\cal I}$) with the system size for varying system sizes from $L =1000$ to $L = 2000$.~(Bottom right panel) Effective central charge,~$c_{\rm eff}$, obtained by fitting to Eq.~\eqref{iEE_sc}. For comparison, the analytical predictions from Eq.~\eqref{ceff} are shown with crosses of the corresponding color. {\it Note the deviation from the predicated central charge value for $K_2< 0.3$. This manifests itself in both the top and bottom right panels. This is due to finite size effects as the isotropic point~($K_2=0$) is approached for the right half of the chain. }}
\end{figure}

Fig.~\ref{XXZ_inter_entropy} shows the EE for different bipartitionings of the system for a total system-size $L = 2000$. The Luttinger parameter for the left half of the chain is fixed to $K_1 = 0.6$, while that for the right half is varied from 0.1 to 0.6 in steps of 0.1. For $K_1 = K_2$, we recover the standard expression for EE~[see Eq.~\eqref{EE_open_chain}] which leads to a central charge of $\simeq0.992$. Note that we use much larger system sizes compared to the Ising chain due to larger finite size effects in the current model. As $K_2$ is lowered, the interface is clearly apparent in the EE~(see left panel). The scaling of the interface EE for different system sizes~[see Eq.~\eqref{iEE_sc}] is shown in the top right panel. The effective central charge obtained from the scaling is plotted in the bottom right panel. Note that as $K_2$ approaches the value 0~(the isotropic point for the right half of the XXZ chain), we start seeing deviations from the predicted central charge value due to the finite size of the system. This is manifest in both the top right and the bottom right panels. It will be interesting to provide an analytical prediction for this finite-size effect. 

\section{Conclusion}
\label{sec:concl}
In this work, we have described the behavior of EE in the CFTs with boundaries and defects. First, we considered the Ising and the free, compactified boson models with Neumann and Dirichlet boundary conditions and computed the change in universal, boundary-dependent contribution to the EE. Next, we computed the behavior of the EE in these models in the presence of conformal defects. In particular, we considered the energy and the duality defects of the Ising model and the interface defect of the free, boson theory. We showed that defects and interfaces, unlike boundaries, manifest themselves in both the leading logarithmic scaling and the ${\cal O}(1)$ term in the EE. 

Here, we concentrated on the von-Neumann entropy as the measure of entanglement. However, entanglement measures like mutual information~\cite{Amico:2007ag} and entanglement spectrum~\cite{Li2008, Haag2012} have seen much use in the characterization of CFTs with boundaries~\cite{Casini:2015woa, Casini:2016fgb, Roy2020a, Mintchev:2020uom} and defects~\cite{Mintchev:2020jhc} for certain geometric arrangements of the subsystem with respect to the boundary or the defect. A particularly simple situation arises for the entanglement Hamiltonian~(${\cal H}_A$) of a subsystem~(A) after bipartioning a CFT with identical boundary conditions on the two ends~\cite{Cardy2016}. Then, ${\cal H}_A$ is related to the CFT Hamiltonian with appropriate boundary conditions~$\alpha, \beta$: 
\begin{align}
\label{ent_ham_formula}
{\cal H}_A = -\frac{1}{2\pi}\ln \frac{e^{-2\pi H_{\alpha\beta}}}{{\rm Tr}\ e^{-2\pi H_{\alpha\beta}}},
\end{align}
where the denominator inside the logarithm originates from the fact that the reduced density matrix $\rho_A$ should be normalized. The above align is to be understood as an equality of the eigenvalues of the two sides the align up to overall shifts and rescalings, which can be absorbed by rescaling the velocity of sound in the corresponding boundary CFT.
The first boundary condition $\alpha$ is inherited from the original system, while the second $\beta$ originates from the entanglement cut and is the free/Neumann boundary condition. Then, the partition function of the CFT on a cylinder with appropriate boundary conditions leads directly to the entanglement Hamiltonian spectrum~\cite{Roy2020a}. 

Consider the case when $\alpha = $ Neumann for the Ising CFT. Then, the corresponding partition function can be written as a sum over characters of the Ising CFT~[see Eq.~\eqref{bs_Ising}]:
\begin{align}
Z_{\rm{NN}}(q) &= {\rm Tr}e^{-2\pi H_{\rm NN}}= \sum_{j=0,\sigma,\epsilon}\Big|\Big\langle \tilde{\frac{1}{16}}\Big|j\Big\rangle\Big|^2 \chi_j(\tilde{q}) = \chi_0(\tilde{q}) + \chi_{\epsilon}(\tilde{q})=\chi_0(q) + \chi_{\epsilon}(q).
\end{align}
Here the parameter $q$ is defined as: \begin{align}
\label{qdef}
q = e^{-2\pi^2/\bar{L}}, \ \tilde{q} = e^{-2\bar{L}},\ \bar{L} = \ln\Big(\frac{2L}{\pi a}\sin\frac{\pi r}{L}\Big),
\end{align}
$r$ is the subsystem size and we have used the explicit form of the modular S-matrix of the Ising CFT~\cite{Cardy1989}. Thus, we find that the partition function gets contribution from two primary fields: $I,\epsilon$. We use the explicit formulas for the characters (see Chapter 8 of Ref.~\cite{diFrancesco1997}):
\begin{align}
\chi_0(q) &= \frac{1}{\eta(q)}\sum_{n\in\mathbb{Z}}\Big[q^{(24n+1)^2/48} - q^{(24n+7)^2/48}\Big],\\\chi_{\epsilon}(q) &= \frac{1}{\eta(q)}\sum_{n\in\mathbb{Z}}\Big[q^{(24n+5)^2/48} - q^{(24n+11)^2/48}\Big],
\end{align}
where $\eta(q)$ is the Dedekind function defined as 
\begin{align}
\label{Dedekind}
\eta(q) = q^{1/24}\varphi(q) = q^{1/24}\prod_{n>0} (1-q^n). 
\end{align}
Expanding in $q$, we get 
\begin{align}
\chi_j(q) &=q^{-1/48+h_j}\sum_{n\geq0}p_j(n)q^n,\ j = 0,\epsilon,
\end{align}
where $p_{0,\epsilon}(i)$ are obtained to be
\begin{align}
p_0(n) &= 1,0,1,1,2,2,3,\ldots,\\ p_\epsilon(n)&=1,1, 1, 1, 2, 2, 3,\ldots.
\end{align}
Thus, the entanglement energies, labeled by two indices: $(j,n)$, are given by
\begin{align}
  \label{eNN}
\varepsilon_{\rm{N}}(j,n) &= -\frac{1}{2\pi}\ln \frac{q^{-1/48 + h_j+n}}{\tilde{q}^{-1/48}\sum\limits_{k=0,\epsilon}\sum\limits_{m\geq0}p_k(m){\tilde{q}}^{h_k+m}}\nonumber\\&=\frac{\bar{L}}{48\pi} +\frac{\pi}{\bar{L}}\Big(-\frac{1}{48}+h_j + n\Big)\nonumber\\&\quad +\frac{1}{2\pi}\ln\sum\limits_{k=0,\epsilon}\sum\limits_{m\geq0}p_k(m)e^{-2\bar{L}(h_k+m)}
\end{align}
with degeneracy at the level $(j,n)$ being given by $p_j(n)$. The lowest entanglement energy level is given by
\begin{align}
\label{e0_tfi_NN}
\varepsilon_{\rm{N}}(0,0) &= \frac{\bar{L}}{48\pi} -\frac{\pi}{48\bar{L}}+\frac{1}{2\pi}\ln\sum\limits_{k=0,\epsilon}\sum\limits_{m\geq0}p_k(m)e^{-2\bar{L}(h_k+m)}.
\end{align}
With respect to this lowest level, the entanglement energies are given by 
\begin{align}
\Delta\varepsilon_{\rm{N}}(j,n) \equiv \varepsilon_{\rm{N}}(j,n) - \varepsilon_{\rm{N}}(0,0) = \frac{\pi}{\bar{L}}\big(h_j + n\big),
\end{align}
and thus, occur at integer (half-integer) values in units of $\pi/\bar{L}$ for $j=0(\epsilon)$. Similar computations can be done for $\alpha = $ Dirichlet~\cite{Roy2020a}. It will be interesting to generalize this computation to the case of CFTs with defects. 

\section{Acknowledgements}
We are particularly grateful to David Rogerson and Frank Pollmann for numerous discussions and collaboration on a related project. AR is supported by a grant from the Simons Foundation (825876, TDN). HS is supported by the ERC Advanced Grant NuQFT.
\bibliography{library_1}
\end{document}